# Investigation of spin orbit torque driven dynamics in ferromagnetic heterostructures

Xinran Zhou[1,*], Hang Chen[2,*], Yu-Sheng Ou[1], Tao Wang[2], Rasoul Barri[2], Harsha Kannan[2], John Q. Xiao[2,†], Matthew F. Doty[1,2,†]

[1]Department of Materials Science and Engineering, University of Delaware, Newark, DE 19716, USA
[2]Department of Physics and Astronomy, University of Delaware, Newark, DE 19716, USA

* These authors contributed equally to this work.
† Corresponding authors: jqx@udel.edu or doty@udel.edu

**Key words:** Time-resolved, magneto-optical Kerr effect, damping-like torque, field-like torque.

## ABSTRACT

We use time-resolved (TR) measurements based on the polar magneto-optical Kerr effect (MOKE) to study the magnetization dynamics excited by spin orbit torques in Py (Permalloy)/Pt and Ta/CoFeB bilayers. The analysis reveals that the field-like (FL) spin orbit torque (SOT) dominates the amplitude of the first oscillation cycle of the magnetization precession and the damping-like (DL) torque determines the final steady-state magnetization. In our bilayer samples, we have extracted the effective fields, $h_{FL}$ and $h_{DL}$, of the two SOTs from the time-resolved magnetization oscillation spectrum. The extracted values are in good agreement with those extracted from time-integrated DCMOKE measurements, suggesting that the SOTs do not change at high frequencies. We also find that the amplitude ratio of the first oscillation to steady state is linearly proportional to the ratio $h_{FL}/h_{DL}$. The first oscillation amplitude is inversely proportional to, whereas the steady state value is independent of, the applied external field along the current direction.

## I. INTRODUCTION

Spin orbit torques (SOTs) have been widely reported to be an efficient way to manipulate or switch the magnetization of an ultrathin ferromagnetic layer (FM), especially in heavy metal (HM)/FM heterostructures [1-3]. SOTs arise from charge-current induced spin current via either the bulk spin Hall effect (SHE) [4-6] in the HM layer or interfacial Rashba effects [7-9]. While the microscopic origins of SOTs are still debatable, in general spin current produces two types of torques on a FM layer, namely, the damping-like (DL) torque and the field-like (FL) torque, which can be described by the Landau–Lifshitz–Gilbert–Slonczewski equation [10,11],

$$\frac{d\boldsymbol{m}}{dt} = -\gamma(\boldsymbol{m} \times \boldsymbol{H}_{eff}) + \alpha\left(\boldsymbol{m} \times \frac{d\boldsymbol{m}}{dt}\right) - \tau_{DL}\boldsymbol{m} \times (\boldsymbol{m} \times \boldsymbol{\sigma}) - \tau_{FL}(\boldsymbol{m} \times \boldsymbol{\sigma})$$

where $\boldsymbol{m}$ is the reduced magnetization vector, $\gamma$ is the Gilbert gyromagnetic ratio, $\alpha$ is the damping factor, $\boldsymbol{\sigma}$ is the polarization direction of the incoming spin (generally in-plane and orthogonal to the current flow direction), $\boldsymbol{H}_{eff}$ is the total effective field that includes the external field $\boldsymbol{H}_{ext}$, anisotropy field $\boldsymbol{H}_{ani}$ and charge current-induced Oersted field $\boldsymbol{h}_{Oe}$, and $\tau_{DL}$ and $\tau_{FL}$ are the coefficients of DL and FL torques, respectively.

When substantial magnetic moments are orthogonal to the vector $\sigma$, $\tau_{DL}$ and $\tau_{FL}$ can be considered as equivalent effective fields $\mathbf{h}_{DL}$ and $\mathbf{h}_{FL}$ in the direction of $\mathbf{m} \times \sigma$ and $\sigma$ with amplitude of $(\hbar/2e)(j\varepsilon/M_s t_{FM})$ and $(\hbar/2e)(j\varepsilon'/M_s t_{FM})$ [12-14], respectively, where $\hbar$ is the reduced Planck's constant, $e$ is the electron charge, $j$ is the charge current density, $M_s$ is the saturation magnetization, $t_{FM}$ is the thickness of the ferromagnetic layer, $\varepsilon$ (half of the spin Hall angle $\theta_{SH}$ of the HM material) and $\varepsilon'$ are the efficiencies of $\mathbf{h}_{DL}$ and $\mathbf{h}_{FL}$, respectively.

There has been great interest in the role of SOTs in magnetization dynamics because of the importance of these dynamics for potential applications of HM/FM heterostructures in magnetic memory devices. Two theoretical models have been proposed to understand the underlying mechanism driving the dynamics: 1) the macro-spin approximation and 2) the micromagnetic model. The macro-spin approximation assumes all microscopic magnetic moments respond identically and coherently to excitations. In the macro-spin model, $\mathbf{h}_{DL}$ induces the rotation of magnetization and $\mathbf{h}_{FL}$ assists the precession of magnetization about the $\sigma$ vector [14,15]. Micromagnetic models are based on the idea that the magnetization reversal is triggered by random nucleation and executed by the domain wall propagation across the magnetic layer. In the micromagnetic model, $\mathbf{h}_{FL}$ triggers the reversal process and $\mathbf{h}_{DL}$ is responsible for driving the domain expansion [16-18].

The DL and FL torques have been widely studied experimentally, mostly through the measurement of quasi-static magneto-resistances [19-22]. Recently we developed the capability to study SOTs in the quasi-DC regime using an all-optical approach known as vector Magneto-optical Kerr effect (MOKE) spectroscopy [23] [24]. These quasi-DC experiments cannot probe the role of the SOTs in magnetization dynamics, in particular the fast oscillations that can be induced by FL torque. Time-resolved MOKE (TRMOKE) [15,25,26] has consequently become a powerful tool to explore magnetization dynamics. However, most of the attention has been focused on the effect of SOTs on switching properties and there has been little work regarding the extraction of SOTs from dynamic measurements. One reason is that the dynamic oscillations of the magnetization are influenced by the magnetization and effective damping of the materials, the total effective fields, and more, which makes the analysis complicated. On the other hand, fitting measured magnetization dynamics with an appropriate theory [27,28] would provide a means of extracting all the relevant parameters. We report TRMOKE experiments that study the evolution of the out-of-plane magnetization component excited by in-plane electric current pulses. With the assistance of microscopic simulation, we demonstrate the extrapolation of the $\mathbf{h}_{DL}$ and $\mathbf{h}_{FL}$ from the experimental TRMOKE spectra. We demonstrate this technique on two samples: 3Py/3Pt and 3Ta/3CoFeB (3 indicates that each layer is 3 nm thick). As a control experiment, we compare our results from values for $\mathbf{h}_{DL}$ and $\mathbf{h}_{FL}$ extracted from DCMOKE measurement of the same samples.

## II. EXPERIMENT
### A. Sample fabrication and characterization

The samples were grown at room temperature on silicon substrates with a thin thermal oxide layer via magnetron sputtering in an argon pressure of $4.5 \times 10^{-3}$ Torr. No heating or external magnetic field was applied to the substrates during sputtering. The devices used for TRMOKE measurements were fabricated by multi-step photolithography. The bilayer sample was first etched into a 20 (length) × 250 (width) μm² area. A coplanar waveguide (CPW) with a signal line having a 10 (length) × 250 (width) μm² opening was then fabricated centered on top of the sample. The CPW was formed by a stack of conductive film with composition 10Ti/600Cu/200Au. We note that the dimensions of this hybrid structure were purposely determined to form a CPW with 50 Ω impedance. To achieve the best impedance matching and high bandwidth transmission of the electric pulse that induces magnetization dynamics, this joint wave guide was placed in series with an external 50 Ω resistor to ground when connected to the electrical current source. The bilayer sample used for DCMOKE measurements was patterned into a 50 × 50 μm$^2$ square where the laser will shine on. Besides, to calibrate the perpendicular $h_{DL}$ field, along current direction two 50-μm-wide metal wires made of 10Ti/80Au were placed on two sides of the sample with a center-to-center distance of 100 μm. The two wires were connected in series when a DC current passes through them, such that the induced out-of-plane Oersted field is doubled and in-plane Oersted fields is cancelled at the center of the bilayer sample, giving rise to a well-defined calibration field.

The static and dynamic magnetic properties of our samples were initially characterized by magnetic hysteresis and ferromagnetic resonance (FMR) measurement, respectively. The magnetic hysteresis loops were measured in a Magnetic Property Measurement System (MPMS) at room temperature. In FMR measurement, the microwave was generated by a two-port vector network analyzer (Agilent N5230C) and guided into samples through a CPW. The dependence of resonant fields and peak widths at half height on microwave frequencies were obtained by sweeping external fields at varying frequencies.

### B. MOKE measurement

MOKE has been a powerful technique for the study of magnetization of magnetic thin films. The interaction between magnetization and light with a well-defined polarization state causes a change of the polarization state that can be quantitatively described by the complex Kerr rotation angle $\theta_{KR}$. Detailed analyses of the $\theta_{KR}$ provides comprehensive information of the magnetization. We recently demonstrated that magnetization re-orientation induced by SOTs in FM/HM bilayers with in-plane magnetization can be fully characterized by polar and quadratic DCMOKE experiments. In the polar MOKE geometry, the polarization of the normally incident light is 45 degrees with respect to the magnetization and the polar Kerr rotation angle $\theta_{KR}^P$ measures the change of out-of-plane magnetization induced by the DL torque. In the quadratic MOKE geometry, the normally-incident light is circularly-polarized and the quadratic Kerr rotation angle $\theta_{KR}^Q$ measures the change of in-plane magnetization induced by the FL torque. Further details about this technique can be found in Refs. [24] and [23]. We also want to note that MOKE-based SOT measurements are immune to thermal effects, providing an advantage relative to electrical measurements such as second harmonic measurement that are impacted by thermal gradients.

The laser used in both DC and TR MOKE experiments is a Ti-sapphire mode-locked pulsed laser with a center wavelength of 780 nm, pulse width of ~100 fs, and a repetition rate of 76 MHz. The polarization control of the normally incident light is achieved by placing either a half-wave plate or a quarter-wave plate before the objective lens which focuses the laser to a spot size < 4 μm on the samples. The $\theta_{KR}$ of the reflected light due to SOTs is measured using a balanced bridge photodiode. In the DCMOKE experiments, a lock-in amplifier (SR 830 DSP) is used to apply an AC current modulated at 1615 Hz to the samples in order to drive the SOTs in the quasi-DC regime. The laser pulses are not correlated to this slow AC current modulation and the balanced bridge photodiode signal is equivalent to that obtained with a continuous wave laser. In the TRMOKE experiments, the AC current source is replaced by a voltage pulse generator (Keysight 81134 A) that is synchronized to the mode-locked Ti: sapphire laser pulses such that the optical probe pulses always arrive at the sample with a well-defined delay time ($\Delta t$) relative to the electrical pump pulse. The voltage pulse generator drives the SOT dynamics in the HM/FM bilayers with pulses that have a temporal width of 6.6 ns and a rising/falling edge of ~50 ps (10% to 90% within 50ps). In the polar MOKE geometry, $\theta_{KR}^P$ of the time-delayed linearly-polarized optical pulses directly measures the instantaneous out-of-plane component induced by SOTs. The temporal evolution of the out-of-plane magnetization induced by the SOTs is revealed by systematically varying the delay time ($\Delta t$). In both TR and DCMOKE experiments, an in-plane external magnetic field ($H_{ext}$) is applied through a pair of Helmholtz coils or an electromagnet.

## III. RESULTS AND DISCUSSION
### A. Static measurement

We illustrate the measurement using the //3Py/3Pt bilayer (// indicates the substrate side). As shown in Fig. 1(a), when a charge current *j* flows along the *x* axis, the spin current with polarization ($\sigma$) in the *y* direction ($j_s = \theta_{SH}\sigma \times j$) [5,29,30] flows into the Py layer due to the bulk SHE in the Pt layer or the interfacial Rashba effect at the HM/FM interface. The magnetization of the Py was initially saturated by a static external magnetic field along the *x* axis. Under this geometry, the DC polar MOKE is proportional to $m_z$, tilted by the effective field $h_{DL}$ of the damping-like torque $T_{DL}$. The in-plane Oersted field $h_{Oe//}$ induced by the current in the HM layer points in the same direction as $h_{FL}$ in the Py /Pt sample (Fig. 1(a)), but points in the direction opposite $h_{FL}$ in the //3Ta/3CoFeB (Fig. 1(b)). Fig. 1(c) and (d) show the normalized magnetic hysteresis loops of the two samples, respectively, obtained by sweeping the magnetic field along the in-plane easy (*y*) and hard (*x*) axes in MPMS. The sharp square loops indicate the existing effective anisotropic fields along the easy axes and the coercivities were found to be 2.6 Oe in //3Py/3Pt and 3.7 Oe in //3Ta/3CoFeB.

For DCMOKE measurements, an electrical current of 10 mA is applied to the //3Py/3Pt sample. We compute using a parallel resistor model that 7 mA flows through Pt layer, corresponding to the current density $j_{Pt} = 4.7 \times 10^{10}$ A/m². The $\theta_{KR}^P$ (in units of balanced bridge photodiode voltage) as a function of $\boldsymbol{H}_{ext}$ is shown in Fig. 2(a) where a hysteresis loop is observed because $h_{DL} \sim m \times \sigma$. We note that we focused the laser spot to a spot size less than 4 um at the center of the 50-um-wide sample such that the induced out-of-plane Oersted field ($h_{Oe\perp}$) from the current was negligible in our experiments. The amplitude of $h_{DL}$ was calibrated by passing current through the calibration wires parallel to the sample (see MOKE measurement section), which introduces $h_{cal}$ in the same direction as $h_{DL}$. The $\theta_{KR}^P$ induced by a fixed calibration field is constant as a function of $\boldsymbol{H}_{ext}$ as shown in Fig. 2 (a) because the calibration field is independent of the magnetization direction. In addition, the $\theta_{KR}^P$ is linearly proportional to the calibration field strength as shown in the inset of Fig. 2 (a). From these calibrations $h_{DL}$ is determined to be 5.23 Oe in the //3Py/3Pt bilayer. The value of $h_{DL}$ corresponds to $\beta_{Pt\_DL}$=8.9 nm where $\beta$ is the SOT efficiency, defined as the ratio of induced effective field to current density in HM layer in unit of Oe/(Am$^{-2}$) or nm.

We next compute the spin Hall angle $\theta_{SH}$ by using the equation $(\hbar/2e)(j\theta_{SH}/2M_s t_{FM}) = h_{DL}$. We note that the effective magnetization $M_{eff}$ is related to $M_s$ and the uniaxial anisotropy field ($H_a$) by the relationship $4\pi M_{eff} = 4\pi M_s - H_a$. $4\pi M_{eff}$ is of order $10^4$ Oe, while $H_a <= 40$ Oe. We therefore make the approximation $M_{eff} \approx M_s$, resulting in $(\hbar/2e)(j\theta_{SH}/2M_{eff} t_{FM}) = h_{DL}$. Using the value of $M_{eff\_Py}$ extracted from the FMR measurements (760 emu/cm³), we calculate a spin Hall angle for the //3Py/3Pt sample of $\theta_{SH} = 0.077$. Similarly, for the //3Ta/3CoFeB sample an electrical current of 10 mA corresponds to $j_{Ta} = 2.1 \times 10^{10}$ A/m² and we determine $\beta_{Ta\_DL}= -10.6$ nm and $\theta_{SH\_Ta} = -0.112$ from the measured data shown in Fig. 2(b). Here, $M_{eff\_CoFeB} = 960$ emu/cm³ was used in the calculation. All the extracted values for these parameters for Pt and Ta are similar to previously reported values [1,31].

To characterize the FL torque, we studied the $\theta_{KR}^Q$ as a function of $H_{ext}$ (red squares) in the Py/Pt and Ta/CoFeB samples as shown in Fig. 2(c) and (d), respectively. The $\theta_{KR}^Q$ changes sign at $H_{ext} = 0$ because the quadratic signal is proportional to the product of $m_x \cdot m_y$ where $m_y$ keeps the same orientation but $m_x$ changes sign when $H_{ext}$ is reversed. The distinctive $1/H$ relation with $H_{ext}$ was observed in our experiment similar to previous reports [23]. To calibrate the $h_{FL}$, we measured the field dependent $\theta_{KR}^Q$ signal from the samples while only a calibration field was applied in the $y$ direction without passing the current through the samples. The calibration field was produced by a 300-μm-wide and 100-nm-thick metallic strip on a Si substrate that is placed underneath the sample. The calibration wire was designed to be wider than the sample width to produce a uniform in-plane magnetic field. A current of 700 mA was passed through the calibration wire, which induced a magnetic field of 2.10 Oe. By taking the linear regression shown in the insets, we determine that Pt and Ta, respectively, produced $h_{FL\_Pt} = 1.56$ Oe and $h_{FL\_Ta} = 0.71$ Oe, which lead to $\beta_{Pt\_FL} = 2.6$ nm in 3Py/3Pt and $\beta_{Ta\_FL} = 2.7$ nm in 3Ta/3CoFeB, respectively.

### B. Dynamic measurement and simulation

We now move beyond the study of SOTs in the quasi-static regime to investigate the SOT-induced magnetization dynamics in the time domain. We perform TRMOKE measurements on a 3Py/3Pt and a 3Ta/3CoFeB sample. The principle of TRMOKE measurements has been discussed in **Section II B**. In Fig. 3(a) we show a schematic diagram of the measurement system. The pulse generator drives the magnetization dynamics with 2 V voltage pulses, corresponding to $j_{Pt} = 2.39 \times 10^{10}$ A/m$^2$ and $j_{Ta} = 1.20 \times 10^{10}$ A/m$^2$, respectively, in the Pt and Ta layers of the two samples.

The induced $\theta_{KR}^P$ as a function of delay time $\Delta t$ for the 3Py/3Pt and 3Ta/3CoFeB samples from $H_{ext} = 50$ Oe to 150 Oe are shown in Fig. 3(b) and (c). The start of the current pulse, schematically depicted by the blue bar at the bottom of Fig. 3(b) and (c), defines $\Delta t = 0$, i.e. when the optical probe pulse arrives simultaneously with the rising edge of the current pulse. In response to the current pulse, $m_z$, which is proportional to the measured voltage (y-axis), undergoes a fast oscillation that decays to a new equilibrium value that depends on both the external magnetic field and the current amplitude. A Fast Fourier Transform (FFT) of the oscillation data (not shown) confirms that there is a single precession frequency. We therefore fit the decaying oscillations with the formula $\sin(2\pi f_0 t + \varphi) \cdot \exp(-t/t_\alpha)$, where $f_0$ is the oscillation frequency, $\varphi$ is the phase of the magnetization precession and $t_\alpha$ is the exponential decay time [32]. In Fig. 4(a), the oscillation frequencies extracted via fits to this functional form (solid data points) are plotted as a function of the applied magnetic field, together with the ferromagnetic resonant frequency (open data points) measured with an FMR spectrometer. We note that the precession frequencies obtained via this fit are in good agreement with those obtained by FFT. The red curve is a fit to the FMR data using the Kittle equation. We find that the dispersion relation extracted from the TRMOKE measurements agrees well with an extension of the FMR fit, indicating that in TRMOKE measurement the magnetization is precessing at FMR frequency. From the fitting curves, the extracted effective magnetizations are $M_{eff\_Py} = 760$ emu/cm$^3$ and $M_{eff\_CoFeB} = 960$ emu/cm$^3$, while the effective anisotropic fields are $H_{ani\_Py} = 31$ Oe and $H_{ani\_CoFeB} = 40$ Oe. The damping of oscillatory TRMOKE signal at each applied field can be calculated using the relationship $\alpha = 1/t_\alpha \gamma (M_s/2 + H)$ for in-plane magnetization in our samples[28]. The calculated data is plotted in Fig. 4(b) and fit by the red curves using the widely applied exponential relation $\alpha(H_{ext}) = \alpha_1 \exp(-H_{ext}/H_0) + \alpha_0$, where $\alpha_1$ is the coefficient, $H_0$ is a constant, and $\alpha_0$ is the effective damping[33]. The effective damping in 3Py/3Pt and 3Ta/3CoFeB samples were found to be 0.024 and 0.008, respectively. They are very close to the damping values $\alpha_{Py/Pt} = 0.022$ and $\alpha_{Ta/CoFeB} = 0.009$ extracted from FMR measurements as shown in the inset of Fig.4 (b). This result indicates the SOT induced magnetization oscillation alternatively can be used to measure effective damping in FM/HM bilayer even at low magnetic field range, while the traditional

FMR measurements typically need to apply a high magnetic field to saturate the film sample in case of the multi domains.

To fully understand the oscillation spectra, we performed micromagnetic simulations using Object Oriented Micromagnetic Framework (OOMMF) software from NIST[34]. The LLG equation was numerically solved using the Oxs_SpinXferEvolve module where both DL and FL terms are included. The constant external magnetic field was applied in the $x$ direction. In the simulation, we used the values of magnetization, anisotropic field, and damping determined from the fits to experimental oscillatory polar MOKE signal. The initial magnetization in our simulation was defined along the $x$ axis while the effective anisotropic field was set along the $y$ axis because even a minimal magnetic field (50 Oe) is strong enough to initialize the magnetization direction [see Figs. 1(c) and (d)]. The computation includes a single current pulse of 6.6 ns duration starting at $t = 0$ ns with 50 ps rising and falling edges, identical to the pulse shape applied in the TRMOKE experiment. The intensity of the current pulse was assigned with the same value as the calculated current density in HM from the experiment. Moreover, a uniform in-plane Oersted field along the $y$ axis ($h_{Oe\_Py} = 0.45$ Oe and $h_{Oe\_CoFeB} = -0.22$ Oe, which are pulse current induced Oersted fields evaluated by the currents flowing through HM metal layer via Ampere's law) was also included in our simulation. The out-of-plane Oersted field was neglected because the laser spot (2 um in diameter) was focused on the sample center position. The spin Hall angles used in our simulation were extracted from the DCMOKE experiment. Both spin polarization vectors $\sigma_{Py}$ and $\sigma_{CoFeB}$ were set in the positive $y$ direction in view of the opposite spin Hall angle signs between the Pt and Ta layers which were, respectively, deposited on the top of the Py and the bottom of the CoFeB layers.

The fit results obtained from the OOMF simulations (red curves) are shown in Fig. 4(c) and match the experimental results very well. This indicates that both $T_{FL}$ and $T_{DL}$ do not change at high frequency. Using the OOMF simulations, we examine the dependence of the plateau value ($P$ in Fig. 4(c)) on the value of $h_{DL}$ and $h_{FL}$. The results (not shown) reveal that P (and thus $\Delta m$) depends only on $h_{DL}$ and not on $h_{FL}$. Therefore, $h_{DL}$ can be extracted from TRMOKE just like the DC case. The extracted values are $h_{DL\_Py/Pt} = 2.62$ Oe and $h_{DL\_Ta/CoFeB} = 1.40$ Oe. With current densities of $j_{Pt} = 2.39 \times 10^{10}$ A/m$^2$ and $j_{Ta} = 1.20 \times 10^{10}$ A/m$^2$, we obtain SOT efficiencies of the DL terms of $\beta = 8.8$ nm and $\beta = -10.2$ nm for the 3Py3Pt and 3Ta/3CoFeB samples, respectively, which are very close to the values in the DC case ($\beta = 8.9$ nm in 3Py3Pt and $\beta = -10.6$ nm in 3Ta/3CoFeB). Through the simulation studies, we also find that the ratio $O/P$ depends linearly on the ratio of $h_{FL}/h_{DL}$, as shown in Fig. 4(d). We find that this linear relationship holds whether $O$ is defined as the amplitude of the first trough (red points) or crest (black points). The same result is found in the Ta/CoFeB system, but is not shown. With this information, the best fit to the oscillation patterns could be quickly determined by changing $h_{FL}$ to match the $O/P$ ratio, leading to the determination of $h_{FL\_Py/Pt} = 0.72$ Oe and $h_{FL\_Ta/CoFeB} = 0.30$ Oe. For the Py/Pt system. The ratio of $(h_{FL}/h_{DL})\_{Py/Pt} = 0.28$ is in good agreement with the ratio of 0.29 determined from the DCMOKE measurement. For the Ta/CoFeB system, the ratio of $(h_{FL}/h_{DL})\_{Ta/CoFeB} = 0.22$ is smaller than the value of 0.26 extracted from the DC MOKE measurement. There are two possible reasons for this discrepancy. First, there is strong interface intermixing at the Ta/CoFeB interface. This has been observed in a few previous experiments and interpreted using a drift-diffusion model [35,36]. This effect may contribute to the SOTs in a manner not captured by our simulation. Second, because current-induced $h_{Oe//}$ partially cancels the $h_{FL}$ due to the layer sequence, the measured magnitude of the oscillation is weak compared with the plateau value seen in Fig. 4 (c). In this case, even a small deviation in $h_{FL}$ would cause a large error on calculation of $h_{FL}/h_{DL}$.

In addition to the linear relationship between the $O/P$ and $h_{FL}/h_{DL}$ ratios, we also found the ratio $O/P$ is inversely proportional to $H_{ext}$, as shown in Fig. 4(e). We demonstrate that $O/P$ is inversely proportional to $H_{ext}$ by showing that the experimental data are well fit by a function of form $a/H+b$ with b values very close to zero. The fit value for coefficient a is loosely related to the Oersted field. The inverse proportionality between the ratio $O/P$ and $H_{ext}$ is also found in our simulation results, which show excellent agreement with the experimental data. We note that the plateau ($P$) is independent of $H_{ext}$. Therefore, the first oscillation amplitude $O$ follows the $1/H_{ext}$ dependence, having the same dependence as $m_y(H_{ext})$ [see Fig. 2(c) and (d)]. This suggests that the initial oscillation in $m_z$ tends to follow $m_y$ and is then suppressed by the large demagnetization field $H_D$, ultimately reaching a final state with a $P$ value that is independent of $H_{ext}$ because $H_{ext}$ is negligible compared with $H_D$.

## IV. CONCLUSIONS

In summary, we use TRMOKE to measure the magnetization dynamics induced by SOTs in two ferromagnetic heterostructures, 3Py/3Pt and 3Ta/3CoFeB. We show that the SOT effective fields $h_{DL}$ can be extracted from the plateau value ($P$) and $h_{FL}$ can be extracted from fitting the TRMOKE spectrum by varying $h_{FL}$ to match the $O/P$ ratio. The extracted effective fields and their efficiencies are in good agreement with those extracted from DCMOKE measurements, indicating that the effective fields do not change at high frequency. These studies provide insight into how $h_{FL}$ and $h_{DL}$ affect the magnetization dynamics. The method to extract $h_{FL}$ and $h_{DL}$ from TRMOKE data may be extendable to other HM/FM bilayer systems.

## AUTHORS CONTRIBUTION AND ACKNOWLEDGEMENTS

J.Q.X and M.F.D designed the experiments; R.B, H.K and H.C fabricated samples; X.Z conducted the TR-MOKE experiments; H.C and Y.S.O performed the FMR and DC MOKE experiments; H.C analyzed the experiments data and performed the simulation; H.C, Y.O., J.Q.X and M.F.D wrote the paper. The work is supported by NSF DMR-1624976 and DOE DE-SC0016380.

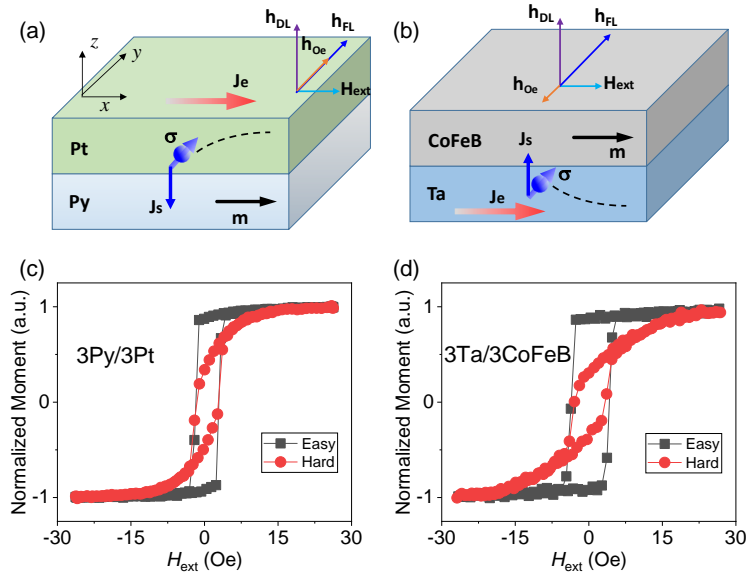

FIG. 1. (a), (b) Schematic of //3Py/3Pt and //3Ta/3CoFeB sample structure and the current-induced effective fields. Magnetization in Py layer is aligned in the same direction with current. Spin current $j_s$ with spin polarization in $y$ direction flowing into Py layer exerts out-of-plane $\boldsymbol{h}_{DL}$ and in-plane $\boldsymbol{h}_{FL}$. $\boldsymbol{h}_{FL}$ and $\boldsymbol{h}_{Oe}$ point opposite directions in two samples due to opposite HM/FM growth sequence and opposite spin Hall angles in Pt and Ta layers. (c) and (d) Magnetic hysteresis loop of //3Py/3Pt and //3Ta/3CoFeB samples, respectively, with external magnetic field along the easy ($y$-) and hard ($x$-) axes.

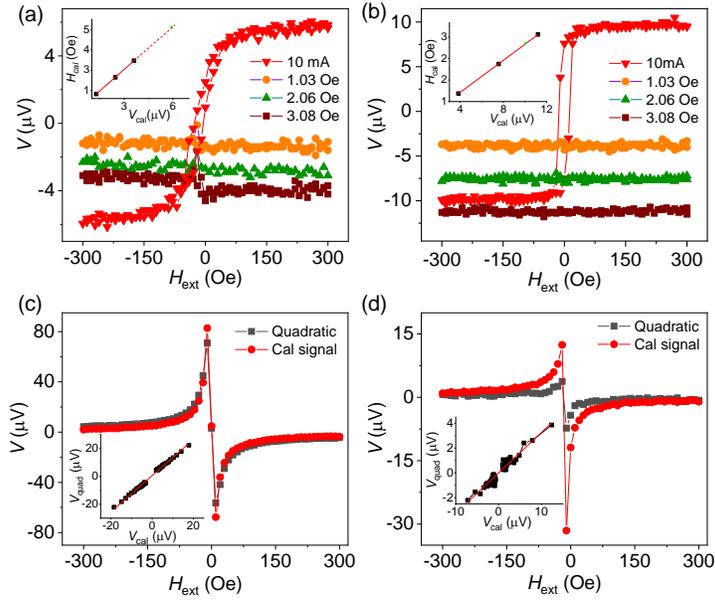

FIG. 2. DC polar MOKE signal as a function of $H_{ext}$ for (a) a 3Py/3Pt and (b) a 3Ta/3CoFeB samples. Red down triangles indicate the MOKE signal when total 10 mA current flows in bilayer structure. $H_{ext}$ was applied along in the same direction as current (*x* direction). The orange spheres, green up triangles and brown squares are the MOKE signals induced solely by perpendicular calibration magnetic fields at 1.03 Oe, 2.06 Oe and 3.08 Oe, respectively. Here we note that the small step-like signals near $H_{ext}$ = 0 in the curve of brown squares is attributed to the misalignment between sample plane and external magnetic field. This effect can be eliminated by averaging the measured data. The inset shows linear relation between calibration field and Measured MOKE signal. The fitting of slope determines the calculation of $h_{DL}$. DC quadratic MOKE measurement for (c) 3Py/3Pt and (d) 3Ta/3CoFeB samples. The black curves were measured MOKE signal solely induced by calibration wires. The insert plots the linear regression of the quadratic MOKE signal induced by current on that of signal induced by calibration fields. $h_{FL}$ has been calibrated from the fitted slope.

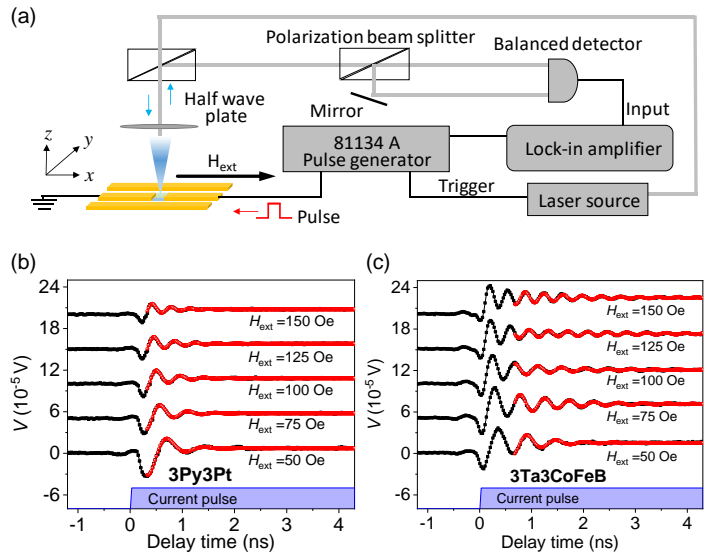

FIG. 3. (a) A schematic view of the TRMOKE measurement setup. Current pulses were sent to sample through a CPW in the $x$–$y$ plane. External magnetic field was applied in the $x$ direction. (b) and (c) The dependence of the measured TRMOKE signal on the delay time for 3Py/3Pt and 3Ta/3CoFeB, respectively. The blue area represents the duration of the current pulse. Measured signal (black squares) at different external magnetic fields from 50 Oe to 150 Oe are offset for clarity. The red curves that overlap the black data points are fit to the experimental data with the formula $\sin(2\pi f_0 t+\varphi)\cdot\exp(-t/t_\alpha)$.

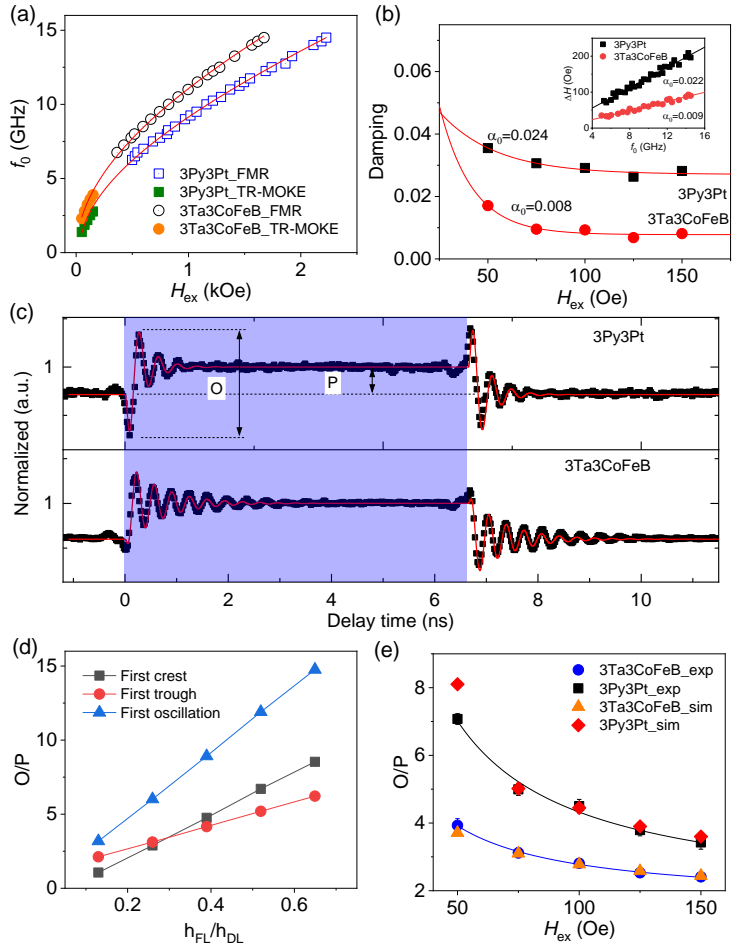

FIG. 4. (a) The oscillation frequencies extracted from TRMOKE measurements (solid data points) and resonant frequencies from FMR measurements (open data points) as a function of $H_{ext}$ for 3Py/3Pt and 3Ta/3CoFeB samples. The red curves are a fit using the Kittle equation. (b) The damping extracted from the TRMOKE data as a function of the applied field. The inset is the linear fit of the FMR linewidth as a function of the frequency using equation $\Delta H = 4\pi f \alpha / \gamma$. (c) Measured TRMOKE response to a current pulse applied over the time window indicated by the blue shaded area for the 3Py/3Pt (top panel) and 3Ta3CoFeB (bottom panel) samples. The solid black squares are experimental data that are fitted (red curves) using micromagnetic OOMMF software. The intensities of the first oscillation cycle and the plateau, i.e. "O" and "P", are indicated by arrows. (d) The extracted ratio $O/P$ as a function of the ratio of $h_{FL}/h_{DL}$ in the simulated Py/Pt system at $H_{ext}$=150 Oe. (e) The dependence of $O/P$ on the external magnetic field in 3Py/3Pt and 3Ta/3CoFeB samples. The black squares and blue spheres represent experimental results. Up and down triangles refer to simulation results. The experimental data were fit using $a/H + b$, where $a = 269.5$, $b = 1.6$ for 3Py/3Pt and $a = 113.9$, $b = 1.6$ for 3Ta/3CoFeB.